\documentclass[a4paper, conference]{IEEEtran}
\IEEEoverridecommandlockouts
\usepackage{cite}
\usepackage{amsmath,amssymb,amsfonts}
\usepackage{algorithmic}
\usepackage{graphicx}
\usepackage{textcomp}
\usepackage{xcolor}

\usepackage{balance}
\usepackage{siunitx}

\usepackage[left=1.57cm, right=1.57cm, top=0.95cm, bottom=2.54cm]{geometry}

\def\BibTeX{{\rm B\kern-.05em{\sc i\kern-.025em b}\kern-.08em
    T\kern-.1667em\lower.7ex\hbox{E}\kern-.125emX}}
\begin{document}

\title{Efficient Circular Polarized Metamaterial\\RF Energy Harvester for Wi-Fi Applications
}

\author{
\IEEEauthorblockN{Stylianos D. Assimonis, and Vincent Fusco}
\IEEEauthorblockA{
    \textit{School of Electron., Electr. Eng. and Comput. Sci.} \\
    \textit{Queen’s University Belfast}\\Belfast, UK\\s.assimonis@qub.ac.uk, v.fusco@ecit.qub.ac.uk}
\and
\IEEEauthorblockN{Manos M. Tentzeris}
\IEEEauthorblockA{
    \textit{School of Electr. and Comput. Eng.} \\
    \textit{Georgia Institute of Technology}\\Atlanta, USA\\etentze@ece.gatech.edu}
}

\maketitle

\begin{abstract}
	A new metamaterial based, circular polarized electromagnetic energy harvester is presented in this work. The structure, which is multilayer and utilises the well-known to the literature geometry of the spiral split ring resonator, operates at the Wi-Fi frequency band. Based on full electromagnetic analysis, the proposed harvester presents high efficiency (i.e., greater than 90\%), adequate angle insensitivity (i.e., 44 deg.) and it is capable of absorbing circular polarised plane waves.
\end{abstract}


\section{Introduction}

Half century ago Veselago theoretically studied media with simultaneously negative values of dielectric permittivity and permeability \cite{Veselago1968}, but only two decades ago Pendry \textit{et al.} managed to build such type of materials, which since then are called \textit{Metamaterials} \cite{Pendry2004}.
The latter structures have reached the attention of many researchers due to their ability to present electric characteristics that are not found in nature, e.g., negative index of refraction, and thus, to manipulate electromagnetic waves, e.g., by preventing propagation, steering or absorbing.

Metamaterial absorbers are capable of perfect absorption of radio frequency (RF) waves   \cite{PhysRevLett.100.207402,Assimonis2019}, and thus, can be used to harvest ambient power and rectify it into DC for the supply of low power consumption electrical devices.
On the other hand, in a typical rectification system, RF power is captured by antennas, and the total harvesting system is usually called \textit{rectenna} \cite{Assimonis2018,ManosApostolos2018}.
Usually, the design of rectenna arrays is a complex procedure, since it depends on the phase of the incident signals \cite{Assimonis2014}: in \cite{Eid2021} Authors proposed the utilisation of a Rotman lens in these RF harvesting systems in order to overcome this problem.
Another solution is the use of metamaterial harvesters (MH): in a typical metamaterial absorber the captured power mainly dissipates in the dielectric and metallic parts of the geometry, however, in a typical MH captured power mainly dissipates in a properly placed in the geometry load, which represents the input impedance of the rectifier of the RF harvesting system \cite{Ramahi2015,Assimonis2015,Zhong2017}.
Most of the works published in the literature on MH present linear polarised structures \cite{Li2020}.
However, since the polarisation of the ambient power is in general not known a priori, circular polarised (CP) MH can offer a solution to this mismatch polarization issue.

The contribution of this work is the design and the numerical analysis in terms of efficiency of a new CP MH, which is highly efficient and adequately angle insensitive.
It utilises a spiral split ring resonator \cite{PhysRevB.69.014402}, which lies on a multilayer structure.
System resonates at Wi-Fi frequency band and it is capable of delivering more than 90\% of the incident power to the rectifier.

\section{Metamaterial Harvester Design}

\begin{figure}[t!]
	\centering
	\includegraphics[width=0.9\linewidth]{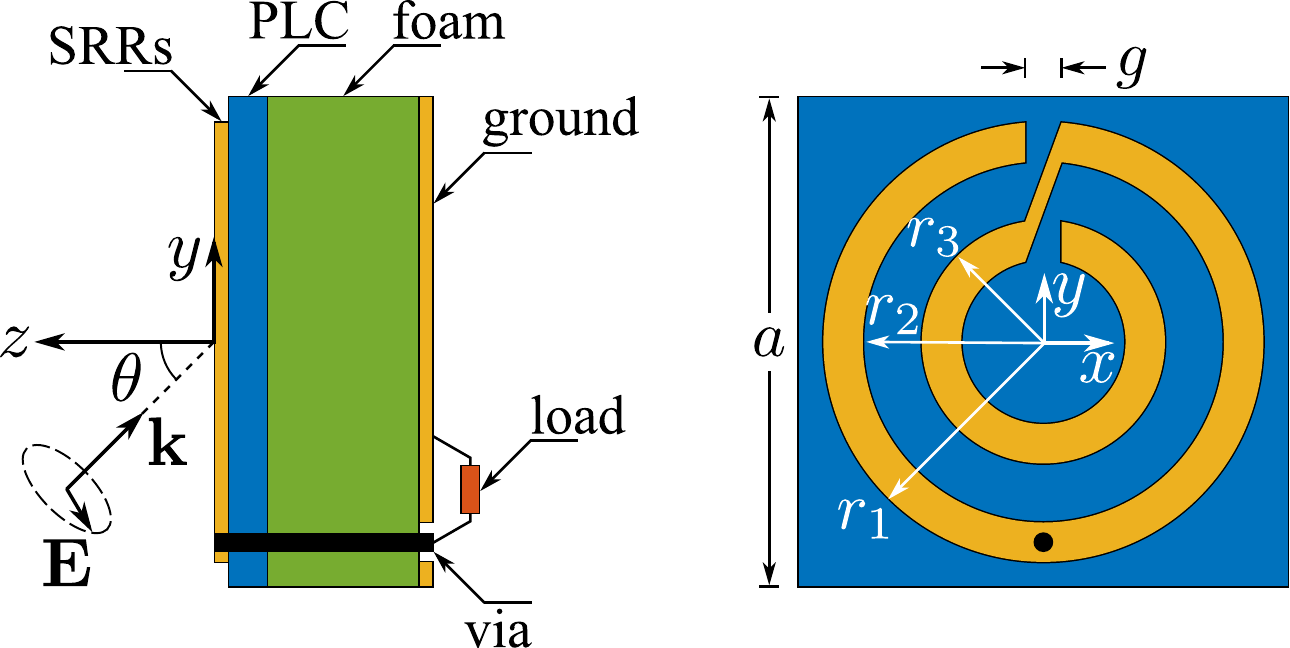}
	\caption{The proposed MH geometry: it is a multilayer structure, which utilises a circular spiral resonator \cite{PhysRevB.69.014402}.}
	\label{fig:geometry}
\end{figure}

\begin{figure*}[t!]
	\centering
	\includegraphics[width=0.990\linewidth]{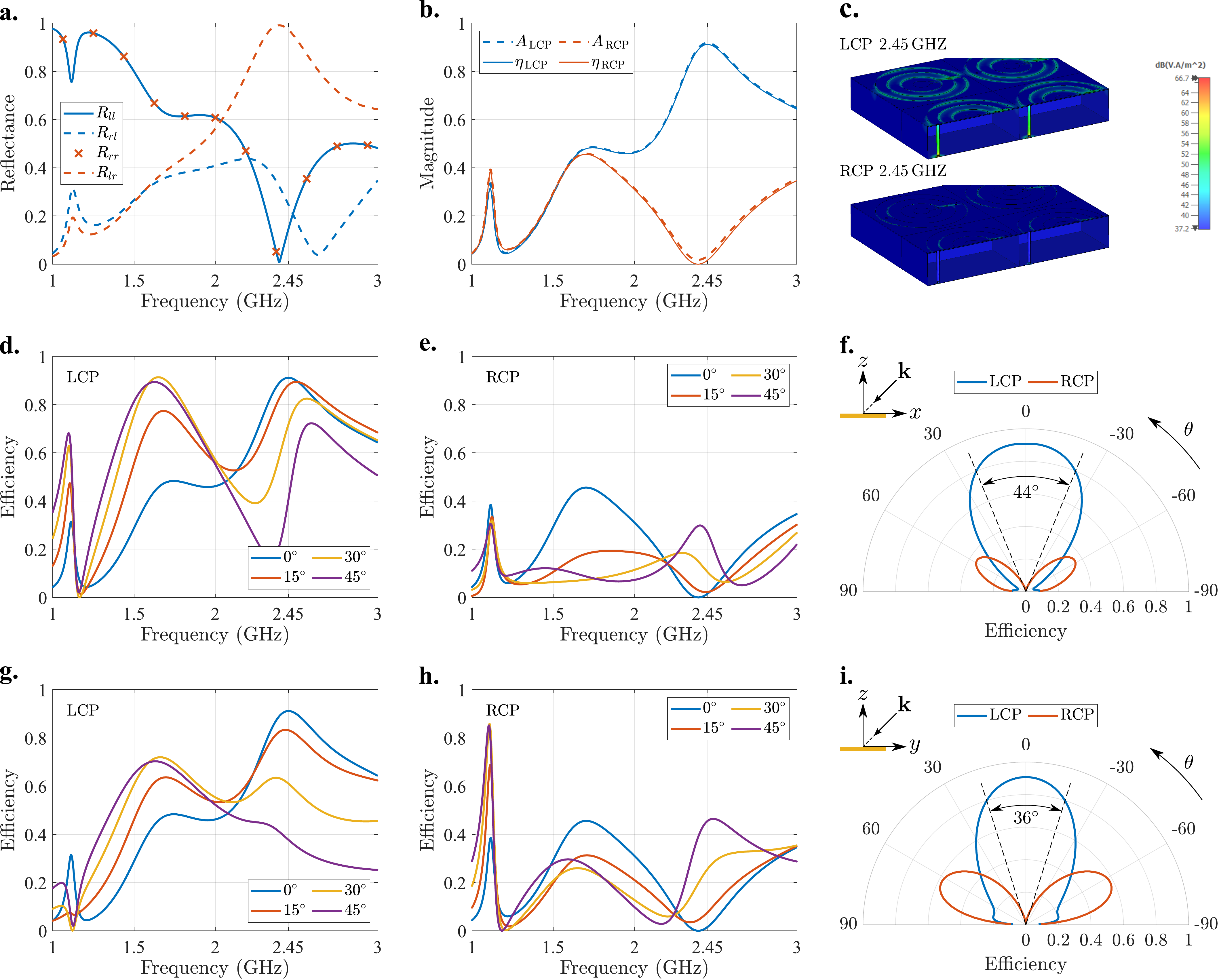}
	\caption{The proposed MH was tested in terms of reflectance \textbf{a.}, absorbance, efficiency \textbf{b.} and power flow \textbf{c.} for normal incidence. It was also tested for oblique incidence. First, the wavenumber $ \mathbf{k} $ was lying in the $ z $-$ x $ plane and efficiency was estimated versus frequency for the LCP case \textbf{d.}, RCP case \textbf{e.}, and versus the angle of incidence $ \theta $ at \SI{2.45}{\GHz}, again for both polarisation cases \textbf{f.} Second, $ \mathbf{k} $ was lying in the $ z $-$ y $ plane and  the efficiency was again estimated versus frequency for the LCP \textbf{g.} and RCP case \textbf{h.}, and versus $ \theta $ \textbf{i.} at \SI{2.45}{\GHz}}
	\label{fig:results}
\end{figure*}

The unit-cell geometry of the proposed MH is depicted in Fig. \ref{fig:geometry}. The multilayer structure consists of two substrates; the top, which is a low-cost flexible copper-clad Liquid crystal polymer (LCP) substrate of thickness \SI{254}{\um} with electric properties of $\epsilon_r=3$ and $\tan\delta=0.003$ \cite{Manos2019}, and the bottom, which is a foam of thickness \SI{10.7}{\mm} and $\epsilon_r=1.01$ and $\tan \delta=0.001$. The latter substrate is grounded. The conductive layer of the metamaterial's pattern lies on the top of the structure and the chosen metal is copper with the thickness of \SI{35}{\um} and the electric conductivity of $5.8\cdot{10}^7$ \SI{}{\siemens\per \meter}. MH pattern is realised via two connected split ring resonators (SRR) forming a circular spiral resonator \cite{PhysRevB.69.014402}.
The dimensions  of the SRRs are
$a=33.31$ mm,
$r_1=14.72$ mm,
$r_2=10.91$ mm,
$r_3=6.51$ mm,
and 
$g=2.02$ mm.
The MH absorbs incident power, which is led into a properly placed load of $50$ $\Omega$ through a vertical via of $0.8$ mm diameter. This load represents the input impedance of a rectification system, which will be connected to the MH, at a later step:
the latter will must be impedance matched to \SI{50}{\ohm} through the utilisation of an impedance matching network.

MH was tested in terms of reflectance $ R $ and absorbance $A$ for normal and oblique incidence of a CP plane  wave versus frequency and angle of incidence $ \theta $ through full electromagnetic analysis using the CST Microwave Studio.
Since the MH is grounded there is no transmission, and thus, absorbance is given by:
\begin{equation}
    A = 1 - \left| R_{\mathrm{++}} \right|^2 - \left| R_{\mathrm{-+}} \right|^2
\end{equation}
where, $R_{\mathrm{++}}$ and $R_{\mathrm{--}}$ are the co- and cross-polarised reflection coefficient, respectively;
in a CP plane wave and for the $R_{\mathrm{++}}$ both incident and reflected signals have the same type of circular polarisation, i.e., both are left- or right-handed circular polarised (LPC or RPC), but for the $R_{\mathrm{-+}}$ incident and reflected signals have different type of circular polarisation.
This property arises from the fact that the handedness of a CP wave changes with the incidence, and specifically, an incident LCP/RCP wave is reflected to a RCP/LCP when impinges to a perfect electric conductor.
%
%
Hence, $A$ was estimated for both polarisation cases, where the incident plane wave is LCP (i.e., $A_{\mathrm{LCP}}$) and RCP (i.e., $A_{\mathrm{RCP}}$).

The simulated results for the reflectance and absorbance are depicted in Fig. \ref{fig:results}a. and \ref{fig:results}b., respectively. For the LCP case it is observed that both corresponding coefficients $ R_{ll} $ (equivalently $ R_{++} $) and $ R_{rl} $ (equivalently $ R_{-+} $) are below $ 0.3 $ at $ 2.45 $ GHz. However, for the RCP case, although the corresponding coefficient $ R_{rr} $ is low and identical to $ R_{ll} $, as expected due to the symmetry of the geometry, power is reflected with a different handedness, and specifically, as LCP, since $ R_{lr} $ is close to unity at $ 2.45 $ GHz.
Indeed, based on the simulated absorbance, incident power is absorbed at $ 2.45 $ GHz for the LCP case, but it is not absorbed for the RCP case, as depicted in Fig. \ref{fig:results}b.

The absorbed power can be dissipated into the dielectric and metallic parts of the MH structure and into the load;
only the latter can be used by the rectification system and can be transformed into DC power.
For this reason, the quantity MH \textit{efficiency} $\eta$ is defined as the ratio of the power, which is consumed by the load over the total incident power, and it is given by:
\begin{equation}
    \eta = \dfrac{P_\mathrm{L}}{P_\mathrm{in}}.
\end{equation}

Efficiency was also tested in terms of handedness of the incident CP plane wave, resulting in $\eta_{\mathrm{LCP}}$ 
and 
$\eta_{\mathrm{RCP}}$, 
and the results are depicted in Fig. \ref{fig:results}b. 
For the LCP case, it can be observed that most of the absorbed power is located into the load rather into the dielectric or metallic parts, since 
$A_{\mathrm{RCP}}$ 
and 
$\eta_{\mathrm{LCP}}$ 
are very close.
Specifically, at $ 2.45 $ GHz the efficiency equals $ 91\% $, while remains over $ 80\% $ for the frequency region of $ 2.32-2.65 $ \SI{}{\GHz}, resulting in full width at half maximum (FWHM) bandwidth of $ 13.3\% $.
For the RCP case, it is evident that the MH does not resonate and cannot capture RF power:  both $ A_{\mathrm{RCP}} $  and $ \eta_{\mathrm{RCP}} $ are very low at \SI{2.45}{\GHz}.

The simulated power density (\SI{}{\watt\per \m^2}) is shown in Fig. \ref{fig:results}c. in dB-scaling.
For the LCP case, the SRRs resonate, power is captured, and it is mainly led to the load (at the end of the via.)
However, for the RCP case, the SRRs do not resonate, and hence, power cannot be captured, as mentioned.

The angle sensitivity with oblique incidence is also tested, since it is important for practical applications of ambient power harvesting, where the location of the ambient power source is not known.
Two cases were examined, where the wavevector $ \mathbf{k} $ confined in the  $z$-$x$ and $z$-$y$ plane, respectively, and the results are depicted in Fig. \ref{fig:results}d. to \ref{fig:results}i.
For both incident cases, power is absorbed only for the LCP case. Also, as the angle of incidence $ \theta $ increases, MH's efficiency decreases at \SI{2.45}{\GHz}, but increases at lower frequencies. Specifically, for RCP, efficiency is over $85\%$ at \SI{1.1}{\GHz} for angle of incidence higher than \SI{30}{\deg}., where normal to the structure components of the electric and magnetic field appear.
At \SI{2.45}{\GHz}, the efficiency of the proposed MH remains over $ 80\% $ for an angle of incidence up to \SI{22}{\deg}. and \SI{18}{\deg}. for the $ z $-$ x $ and $ z $-$ y $ case, respectively, resulting in beamwidth of \SI{44}{\deg}. and \SI{36}{\deg}., respectively.
Thus, the proposed MH presents an adequate angle insensitivity.

\balance

\section{Conclusion}

In this work a new circular polarised MH was designed and numerically analysed through full electromagnetic analysis. The geometry operates at Wi-Fi frequency band and presents high efficiency and adequate angle insensitivity. Thus, the proposed MH is an attractive RF front-end solution for the design of high efficiency and low-complexity, circular polarised RF energy harvesters. By further exploiting this geometry, the proposed design will be fabricated and measured.


\begin{thebibliography}{10}
	\providecommand{\url}[1]{#1}
	\csname url@samestyle\endcsname
	\providecommand{\newblock}{\relax}
	\providecommand{\bibinfo}[2]{#2}
	\providecommand{\BIBentrySTDinterwordspacing}{\spaceskip=0pt\relax}
	\providecommand{\BIBentryALTinterwordstretchfactor}{4}
	\providecommand{\BIBentryALTinterwordspacing}{\spaceskip=\fontdimen2\font plus
		\BIBentryALTinterwordstretchfactor\fontdimen3\font minus
		\fontdimen4\font\relax}
	\providecommand{\BIBforeignlanguage}[2]{{%
			\expandafter\ifx\csname l@#1\endcsname\relax
			\typeout{** WARNING: IEEEtran.bst: No hyphenation pattern has been}%
			\typeout{** loaded for the language `#1'. Using the pattern for}%
			\typeout{** the default language instead.}%
			\else
			\language=\csname l@#1\endcsname
			\fi
			#2}}
	\providecommand{\BIBdecl}{\relax}
	\BIBdecl
	
	\bibitem{Veselago1968}
	V.~G. Veselago, ``The electrodynamics of substances with simultaneously
	negative values of $\epsilon$ and $\mu$,'' \emph{Soviet Physics Uspekhi},
	vol.~10, no.~4, pp. 509--514, Jan.-Feb. 1968.
	
	\bibitem{Pendry2004}
	D.~R. Smith, J.~B. Pendry, and M.~C.~K. Wiltshire, ``Metamaterials and negative
	refractive index,'' \emph{Science}, vol. 305, no. 5685, pp. 788--792, 2004.
	
	\bibitem{PhysRevLett.100.207402}
	N.~I. Landy, S.~Sajuyigbe, J.~J. Mock, D.~R. Smith, and W.~J. Padilla,
	``Perfect metamaterial absorber,'' \emph{Phys. Rev. Lett.}, vol. 100, p.
	207402, May 2008.
	
	\bibitem{Assimonis2019}
	S.~D. Assimonis and V.~Fusco, ``Polarization insensitive, wide-angle,
	ultra-wideband, flexible, resistively loaded, electromagnetic metamaterial
	absorber using conventional inkjet-printing technology,'' \emph{Scientific
		Reports}, vol.~9, no.~1, p. 12334, Aug 2019.
	
	\bibitem{Assimonis2018}
	S.~D. Assimonis, V.~Fusco, A.~Georgiadis, and T.~Samaras, ``Efficient and
	sensitive electrically small rectenna for ultra-low power rf energy
	harvesting,'' \emph{Scientific Reports}, vol.~8, no.~1, p. 15038, Oct 2018.
	
	\bibitem{ManosApostolos2018}
	V.~Palazzi, J.~Hester, J.~Bito, F.~Alimenti, C.~Kalialakis, A.~Collado,
	P.~Mezzanotte, A.~Georgiadis, L.~Roselli, and M.~M. Tentzeris, ``A novel
	ultra-lightweight multiband rectenna on paper for rf energy harvesting in the
	next generation lte bands,'' \emph{IEEE Transactions on Microwave Theory and
		Techniques}, vol.~66, no.~1, pp. 366--379, 2018.
	
	\bibitem{Assimonis2014}
	S.~D. Assimonis, S.-N. Daskalakis, and A.~Bletsas, ``Efficient rf harvesting
	for low-power input with low-cost lossy substrate rectenna grid,'' in
	\emph{2014 IEEE RFID Technology and Applications Conference (RFID-TA)}, 2014,
	pp. 1--6.
	
	\bibitem{Eid2021}
	A.~Eid, J.~G.~D. Hester, and M.~M. Tentzeris, ``5g as a wireless power grid,''
	\emph{Scientific Reports}, vol.~11, no.~1, p. 636, Jan 2021.
	
	\bibitem{Ramahi2015}
	T.~S. Almoneef and O.~M. Ramahi, ``Metamaterial electromagnetic energy
	harvester with near unity efficiency,'' \emph{Applied Physics Letters}, vol.
	106, no.~15, p. 153902, 2015.
	
	\bibitem{Assimonis2015}
	S.~Assimonis, T.~Kollatou, D.~Tsiamitros, D.~Stimoniaris, T.~Samaras, and
	J.~Sahalos, ``High efficiency and triple-band metamaterial electromagnetic
	energy hervester,'' in \emph{2015 9th International Conference on Electrical
		and Electronics Engineering (ELECO)}, 2015, pp. 320--323.
	
	\bibitem{Zhong2017}
	H.-T. Zhong, X.-X. Yang, X.-T. Song, Z.-Y. Guo, and F.~Yu, ``Wideband
	metamaterial array with polarization-independent and wide incident angle for
	harvesting ambient electromagnetic energy and wireless power transfer,''
	\emph{Applied Physics Letters}, vol. 111, no.~21, p. 213902, 2017.
	
	\bibitem{Li2020}
	L.~Li, X.~Zhang, C.~Song, and Y.~Huang, ``Progress, challenges, and perspective
	on metasurfaces for ambient radio frequency energy harvesting,''
	\emph{Applied Physics Letters}, vol. 116, no.~6, p. 060501, 2020.
	
	\bibitem{PhysRevB.69.014402}
	J.~D. Baena, R.~Marqu\'es, F.~Medina, and J.~Martel, ``Artificial magnetic
	metamaterial design by using spiral resonators,'' \emph{Phys. Rev. B},
	vol.~69, p. 014402, Jan 2004.
	
	\bibitem{Manos2019}
	A.~Eid, J.~Hester, and M.~M. Tentzeris, ``A scalable high-gain and
	large-beamwidth mm-wave harvesting approach for 5g-powered iot,'' in
	\emph{2019 IEEE MTT-S International Microwave Symposium (IMS)}, 2019, pp.
	1309--1312.
	
\end{thebibliography}
\end{document}